\documentclass[preprint]{aastex}
\usepackage{hyperref}
\usepackage{aas_macros}
\usepackage{natbib}
\usepackage{amsmath}
\usepackage{amsfonts}
\usepackage{amssymb}

\shortauthors{G.V. Vereshchagin}
\shorttitle{Cosmic horizon for GeV sources and photon-photon scattering}

\begin{document}

\title{Cosmic horizon for GeV sources and photon-photon scattering}
\author{G.~V.~Vereshchagin\altaffilmark{1,2,3}}
\affil
{$^{1}$ ICRANet, p.le della Repubblica, 10, 65122 Pescara, Italy and \\
$^{2}$ ICRA and Deptartment of Physics, University of Rome ``Sapienza'', \\ p.le A. Moro 5, 00185 Rome, Italy and \\
$^{3}$ ICRANet-Minsk, National Academy of Sciences of Belarus, \\ Nezavisimosti av. 68, 220072 Minsk, Belarus}

\begin{abstract}
Photon-photon scattering of gamma-rays on the cosmic microwave background has been
studied using the low energy approximation of
the total cross section
by \citet{1989ApJ...344..551Z,1990ApJ...349..415S}. Here,  the cosmic
horizon due to photon-photon scattering is accurately determined  using the exact cross section and we find
that photon-photon scattering dominates over the pair production at energies
smaller than 1.68 GeV and at redshifts larger than 180.
\end{abstract}

\keywords{Cosmic rays; Background radiation.}

%\ccode{PACS numbers: 98.70.Sa, 98.70.Vc}

\section{Introduction}

Photon-photon scattering $\gamma_{1}\gamma_{2}\longrightarrow\gamma
_{1}^{\prime}\gamma_{2}^{\prime}$\ is a nonlinear electrodynamical process,
allowed by quantum electrodynamics, but much less well known compared to pair
production from two photons, $\gamma_{1}\gamma_{2}\longrightarrow e^{+}e^{-}$. The latter has not yet been directly observed, but has well known
astrophysical implications \citep{1992ApJ...390L..49S,2008A&A...487..837F,2010PhR...487....1R}. The total cross section
of photon-photon scattering in the low energy approximation can be found
in most textbooks on the topic, see e.g. \cite{1982els..book.....B}. The exact
cross section for arbitrary energies has been determined numerically, see e.g.
\cite{1951PhRv...83..776K,1965NCim...35.1182D}. %Recently experimental verification of this process has been proposed \citep{Drebot2016}.

Photon-photon scattering involving cosmic microwave background (CMB) photons has been considered in the cosmological context in
\citet{1989ApJ...344..551Z,1990ApJ...349..415S}. Using the low energy
approximation these authors obtained analytical expressions for the cosmic
horizon, i.e., the redshift as a function of particle energy found by equating the
optical depth to unity. In the limit of large redshifts in the
Einstein-de-Sitter universe they found
\begin{equation}
z=5.002\times10^{3}T_{2.7}^{-4/5}h_{50}^{2/15}\varepsilon_{obs}^{-2/5}%
,\label{EHslope}%
\end{equation}
where the dimensionless observed energy $\varepsilon_{obs}=E_{obs}/(m_{e}c^{2})$ of the gamma-ray photon is expressed in terms of the electron rest mass energy $m_{e}c^{2}$,  the temperature $T$ of the cosmic microwave background is normalized to $2.7$ K, and the Hubble parameter is $H_{0}=50 h_{50}$ $km/s/Mpc$.

It was recognized that the slope of the relation (\ref{EHslope}) differs slightly
from the slope of the Fazio-Stecker relation
\citep{1970Natur.226..135F}, corrected by \cite{1989ApJ...344..551Z} to read
\begin{equation}
z=8.84\times10^{3}\varepsilon_{obs}^{-0.478}.\label{BWslope}%
\end{equation}
The horizon relations for pair production from two photons $\gamma_{1}\gamma
_{2}\longrightarrow e^{+}e^{-}$ and for the photon-photon scattering
$\gamma_{1}\gamma_{2}\longrightarrow\gamma_{1}^{\prime}\gamma_{2}^{\prime}$
were determined, and found to have a crossing point at the approximate redshift
$z_{cr}\simeq3\times10^{2}$. The authors concluded that photon-photon
scattering dominates over pair production at larger redshifts.

In this paper we revisit the derivation of the cosmic horizon relation for photon-photon scattering on the CMB photon background by
considering the exact cross section found by
\cite{1998PhRvD..57.2443D}, instead of the approximate one valid only in the low
energy limit. One might argue that the difference between the exact cross
section and its low energy approximation would be small even near the pair
production threshold, but in fact the ratio between the exact and approximate cross
sections at the threshold is $7.26$. We emphasize that due to the very similar slopes
of the two functions (\ref{EHslope}) and (\ref{BWslope}), even a small change in the
cross section results in a significant shift of the crossing point $z_{cr}$. In
this paper it is shown that the above mentioned crossing point is located at
a lower redshift than previously determined, namely $z_{cr}\simeq180$, and the
corresponding photon energy is $1.68$ GeV.

These new results are essential for photon propagation from sources located at very high redshifts, above 100. Specifically, such photons are present in models involving exotic particles, which decay into photons in a high redshift universe, see e.g. \cite{2006MNRAS.369.1719M,2017JCAP...03..043P} and references therein.

\section{Exact cross section for photon-photon scattering}

The approximate cross section for photon-photon scattering in the low energy
approximation is given by%
\begin{equation}
\sigma=\frac{7\times139}{3^{4}5^{3}\pi}\alpha^{4}r_{0}^{2}\varepsilon_{CM}%
^{6}, \label{sigmaEHan}%
\end{equation}
where $\alpha$ is the fine structure constant, $r_0$ is classical electron radius,
$\varepsilon_{CM}=\sqrt{\varepsilon_{1}\varepsilon_{2}\left(
1-\cos\vartheta\right)  /2}=\sqrt{x\left(  1-\cos\vartheta\right)  /2}$ is the
center-of-momentum energy, $\varepsilon_{1}=h\nu_1/m_e c^2$ and $\varepsilon_{2}=h\nu_2/m_e c^2$ are, respectively, the
dimensionless energies of the high energy photon and the CMB photon, $h$ is Planck's constant, $\nu$ is the photon frequency, $m_e$ the electron mass and $c$ the speed of light. The cosmic horizon is obtained taking
into account both the cosmic evolution of the CMB and the cosmological redshift of
the high energy photon as follows. First, the cross section is averaged over all
angles and integrated over the photon energy using the isotropic distribution function for
the CMB photons. Then the result is integrated over
distance (redshift) to obtain the optical depth as a function of the redshift of
the source and the energy of the observed photon. Equating this optical depth to
unity results in the relation between the redshift of the source and the observed
energy. Equation (\ref{EHslope}) was obtained precisely in this way.

\begin{figure}[ht]%
\centering
\includegraphics[
width=.9\textwidth
]%
{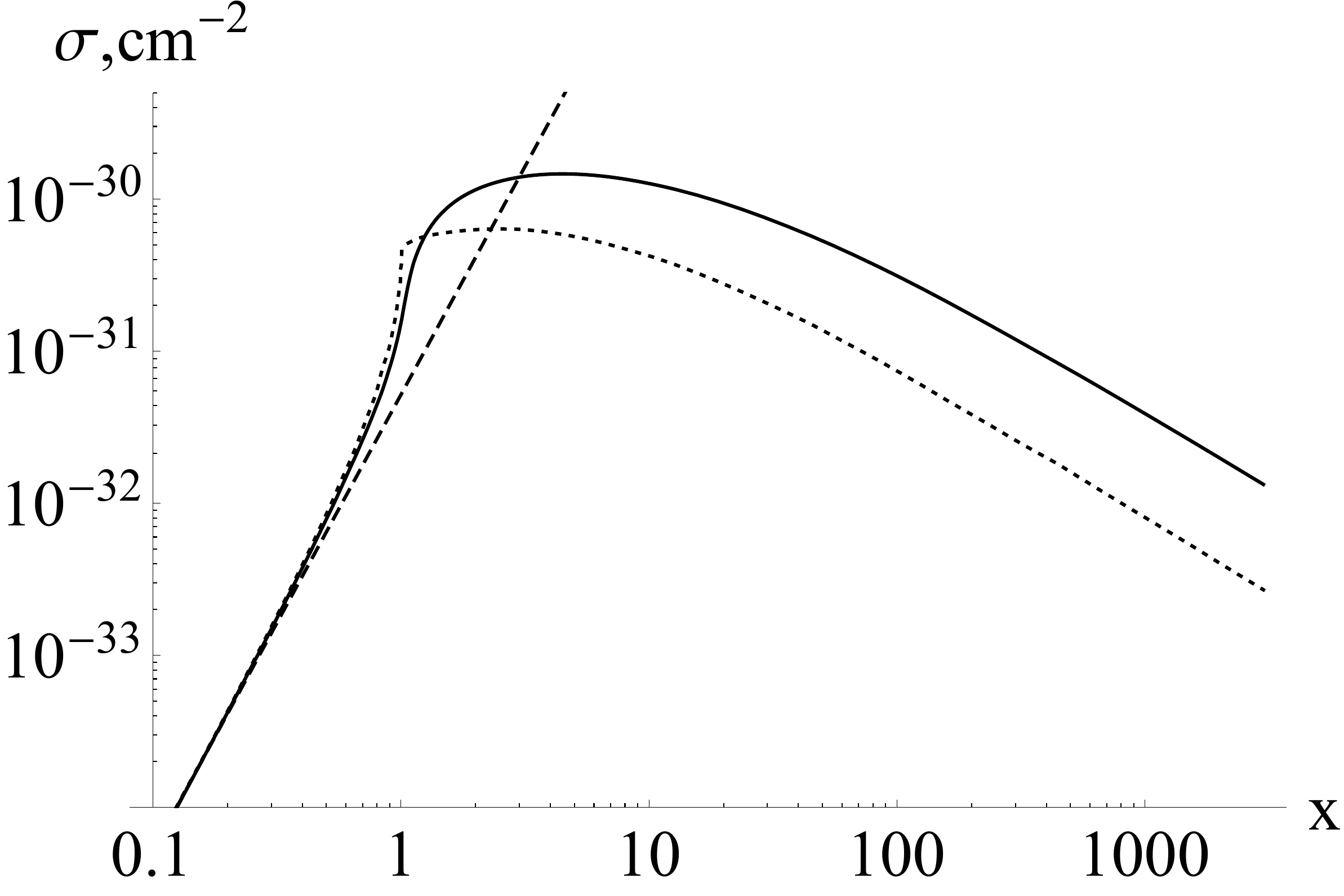}%
\caption{Total cross section of the photon-photon scattering as a function of
the variable $x\equiv\varepsilon_{1}\varepsilon_{2}$ for head on collisions with $\vartheta=\pi$\ (dotted curve).
The solid curve shows the angle averaged cross section, see Eq.~(\ref{tau}).
The dashed line shows the low energy approximation $\sim x^{3}$.}%
\label{sigmaEHfig}%
\end{figure}

Instead of using the approximate cross section (\ref{sigmaEHan}), we take the
exact cross section represented by the dotted curve in Fig.~\ref{sigmaEHfig}.%
%TCIMACRO{\FRAME{ftbpFU}{3.1733in}{2.1208in}{0pt}{\Qcb{Total cross section of
%the photon-photon scattering as a function of the variable $x$ for head on
%collisions with $\vartheta=\pi$\ (dashed curve). Solid curve shows the angle
%averaged cross section, see eq. (\ref{tau}). Dashed line shows the low energy
%approximation $\sim x^{3}$.}}{\Qlb{sigmaEHfig}}{ehcrosssection.eps}%
%width 3.1733in;  height 2.1208in;  depth 0pt;  original-width 5.2424in;
%original-height 3.4888in;  cropleft "0";  croptop "1";  cropright "1";
%cropbottom "0";  filename 'EHcrosssection.eps';file-properties "XNPEU";}}}%
%BeginExpansion
%EndExpansion
It is important to emphasize that the exact form of the cross section near the
threshold for pair production at $x\equiv\varepsilon_{1}\varepsilon_{2}=1$ is crucial. The solid curve in Fig.~\ref{sigmaEHfig}
represents the angle averaged cross section. This
function is integrated further with the photon distribution function and it is
the value near its peak which determines the dependence of the optical depth
on the particle energy and distance. It is clear that averaging over angles
makes the cross section smoother and shifts the peak to higher values of
the variable $x$.

\section{The optical depth and the cosmic horizon for photon-photon
scattering}

The computation of the optical depth is straightforward, for details see e.g.
\citet{2016Ap&SS.361...82R}. The optical depth is given by%
\begin{align}
\tau &  =4\pi\frac{c}{H_{0}}\left(  \frac{h}{m_{e}c}\right)  ^{-3}\left(
\frac{kT_{0}}{m_{e}c^{2}}\right)  ^{3}\left(\frac{1}{y_0}\right)^3\int_{0}^{z}\frac{dz^{\prime}}{\left(
1+z^{\prime}\right)  ^{4}H\left(  z^{\prime}\right)  }\times\label{tau}\\
&  \int_{0}^{\infty}\frac{x^{2}dx}{\exp\left(  x/y\right)  -1}\int_{0}^{\pi
}\sigma\left(  x,y,z^{\prime},\vartheta\right)  \left(  1-\cos\vartheta
\right)  \sin\vartheta d\vartheta,\nonumber
\end{align}
where the variables $x=\varepsilon_{1}\varepsilon_{2}$ and $y=\varepsilon_{2}kT/(m_{e}c^{2})$ depend on the redshift, the index ``0" refers to the observed photon at redshift $z=0$,
\begin{equation}
H(z)=[\Omega_{r}(1+z)^{4}+\Omega_{M}(1+z)^{3}+\Omega_{\Lambda}]^{1/2}%
,\label{free}%
\end{equation}
and $\Omega_{r}=8.4\times10^{-5}$, $\Omega_{M}=0.3089$ and $\Omega_{\Lambda}=0.6911$\ are the present normalized densities
of radiation, matter and dark energy, respectively, while $H_0=67.7$ km/s/Mpc. We compute the integral
(\ref{tau}) numerically, using the latest cosmological parameters given by the \cite{2016A&A...594A..13P}.

\begin{figure}[ht]%
\centering
\includegraphics[
width=0.9\textwidth
]%
{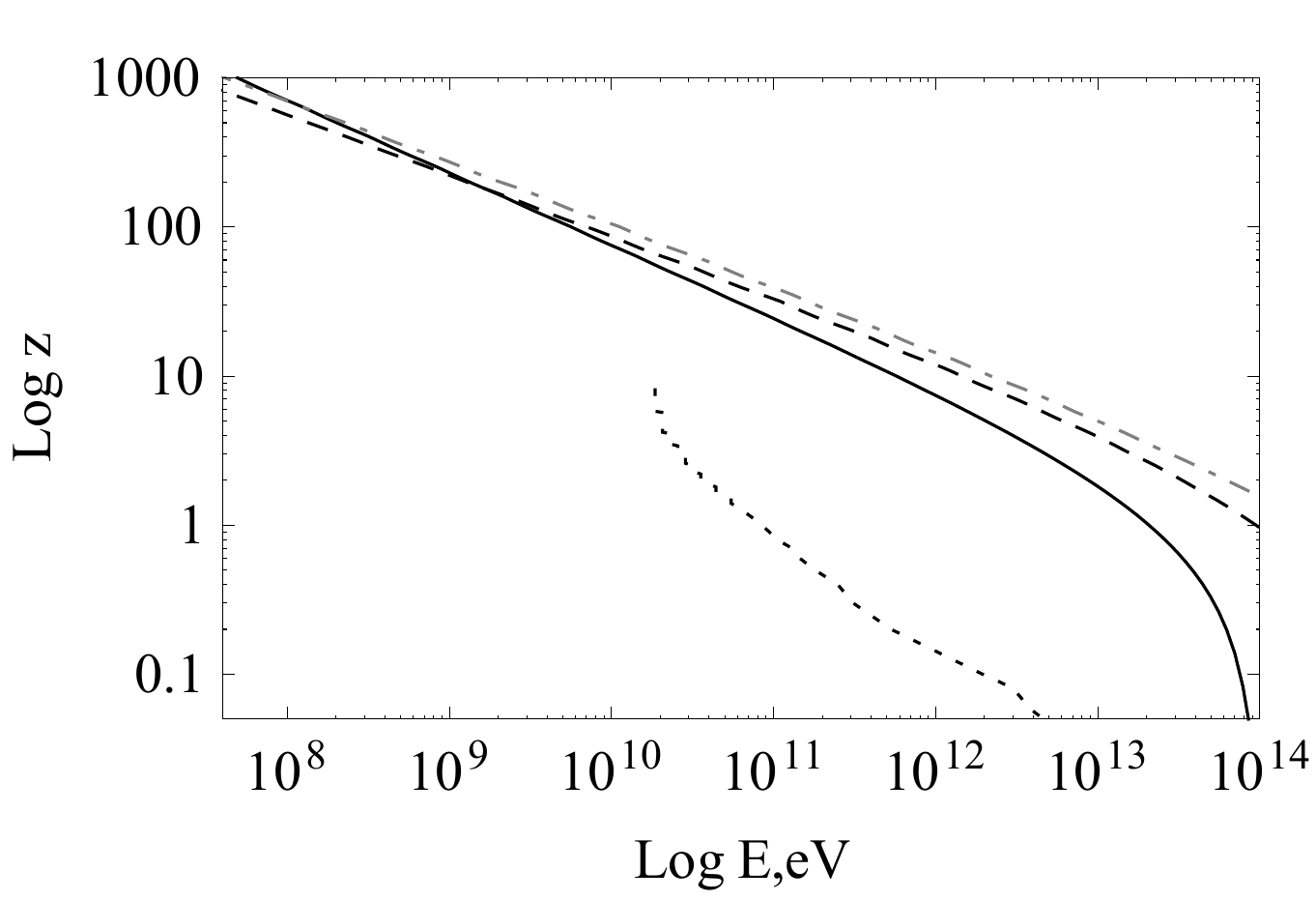}%
\caption{Cosmic horizon (defined from the condition $\tau=1$) for photon-photon scattering (dashed curve) and for pair production from two photons (solid curve) as function of energy $E=h\nu_1$ of the high energy photon measured today on Earth. The dotted curve is the cosmic horizon due to the extragalactic background light, taken from \protect\citet{2013ApJ...768..197I}. For comparison also the condition $\tau=5$ for photon-photon scattering is shown by dash-dotted curve.}%
\label{zefig}%
\end{figure}

The result for optical depth $\tau=1$ is shown in Fig. \ref{zefig} by dashed curve as a function of the energy $E=h\nu_1$ of the high energy photon observed today on Earth. Also shown is the cosmic horizon for the pair production from
two photons, computed in \cite{2016Ap&SS.361...82R}.
%TCIMACRO{\FRAME{ftbpFU}{3.1096in}{1.8615in}{0pt}{\Qcb{Cosmic horizon for
%photon-photon scattering and for pair production from two photons as function
%of particle energy $y_{0}=E/E_{BW}$.}}{\Qlb{zefig}}{zerelation.eps}%
%width 3.1096in;  height 1.8615in;  depth 0pt;  original-width 3.8522in;
%original-height 2.2905in;  cropleft "0";  croptop "1";  cropright "1";
%cropbottom "0";  filename 'zerelation.eps';file-properties "XNPEU";}}}%
%BeginExpansion
%EndExpansion
The high redshift (low energy) asymptotes for cosmic horizons shown in Fig. \ref{zefig} are power
laws given approximately by
\begin{equation}
z=0.786\left(\frac{E}{E_{BW}}\right)^{-0.405}, \label{zeEH}%
\end{equation}
for photon-photon scattering and by%
\begin{equation}
z=0.257\left(\frac{E}{E_{BW}}\right)^{-0.488}, \label{zeBW}%
\end{equation}
for pair production from two photons, respectively, where $E_{BW}=(m_{e}c^{2})^{2}/kT_{0}\simeq1.11\times10^{15}$ eV.

The photon-photon scattering starts to dominate over pair production at
energies smaller than $E_{cr}=1.68$ GeV and redshifts larger than
$z_{cr}\simeq180$. It is important to underline that, unlike pair production
by two photons, photon-photon scattering of a high energy photon on low energy
background is a process that ``splits" the high energy photon into two photons, each of
which carries away on average half of the initial energy. This makes the mean free path shown in 
Fig.~\ref{zefig} also equivalent to the energy loss distance.
For the sake of comparison also the result corresponding to $\tau=5$ is shown by dash-dotted curve.

In Fig.~\ref{zefig} the cosmic horizon due to extragalactic background light (EBL) is represented as the dotted curve. It is clear that the dominance of the photon-photon scattering occurs at energies lower than those relevant for the absorption by the EBL, and at much larger redshifts.

\section{Conclusions}

The photon-photon scattering at cosmological distances is revisited
using the recently obtained exact cross section rather than the low
energy approximation adopted in previous work. Since the exact cross section
near the pair production threshold is larger than the approximate one
obtained in the low energy limit, the dominance of the photon-photon
scattering over pair production by two photons occurs at smaller redshifts
than previously thought, namely redshifts larger than $z_{cr}\simeq180$. This
corresponds to energies smaller than $E_{cr}=1.68$ GeV. These results are relevant for high energy photons produced during the Dark Ages which follows the decoupling of matter and radiation, e.g. by photons resulting from the decay of unstable particles.

\smallskip

{\bf Acknowledgements.}
I would like to thank the anonymous referee for his/her remarks and suggestions which improved presentation of this paper.

%\bibliographystyle{mn2e}
%\bibliography{GZKnu}

\begin{thebibliography}{99}

\bibitem[\protect\citeauthoryear{{Berestetskii}, {Lifshitz} \&
  {Pitaevskii}}{{Berestetskii} et~al.}{1982}]{1982els..book.....B}
{Berestetskii} V.~B.,  {Lifshitz} E.~M.,    {Pitaevskii} V.~B.,  1982, {Quantum
  Electrodynamics}.
Elsevier

\bibitem[\protect\citeauthoryear{{De Tollis}}{{De
  Tollis}}{1965}]{1965NCim...35.1182D}
{De Tollis} B.,  1965, Il Nuovo Cimento, 35, 1182

\bibitem[\protect\citeauthoryear{{Dicus}, {Kao} \& {Repko}}{{Dicus}
  et~al.}{1998}]{1998PhRvD..57.2443D}
{Dicus} D.~A.,  {Kao} C.,    {Repko} W.~W.,  1998, \prd, 57, 2443

\bibitem[\protect\citeauthoryear{{Fazio} \& {Stecker}}{{Fazio} \&
  {Stecker}}{1970}]{1970Natur.226..135F}
{Fazio} G.~G.,  {Stecker} F.~W.,  1970, \nat, 226, 135

\bibitem[\protect\citeauthoryear{{Franceschini}, {Rodighiero} \&
  {Vaccari}}{{Franceschini} et~al.}{2008}]{2008A&A...487..837F}
{Franceschini} A.,  {Rodighiero} G.,    {Vaccari} M.,  2008, \aap, 487, 837

\bibitem[\protect\citeauthoryear{{Inoue}, {Inoue}, {Kobayashi}, {Makiya},
  {Niino} \& {Totani}}{{Inoue} et~al.}{2013}]{2013ApJ...768..197I}
{Inoue} Y.,  {Inoue} S.,  {Kobayashi} M.~A.~R.,  {Makiya} R.,  {Niino} Y.,
  {Totani} T.,  2013, \apj, 768, 197

\bibitem[\protect\citeauthoryear{{Karplus} \& {Neuman}}{{Karplus} \&
  {Neuman}}{1951}]{1951PhRv...83..776K}
{Karplus} R.,  {Neuman} M.,  1951, Physical Review, 83, 776

\bibitem[\protect\citeauthoryear{{Mapelli}, {Ferrara} \& {Pierpaoli}}{{Mapelli}
  et~al.}{2006}]{2006MNRAS.369.1719M}
{Mapelli} M.,  {Ferrara} A.,    {Pierpaoli} E.,  2006, \mnras, 369, 1719

\bibitem[\protect\citeauthoryear{{Planck Collaboration}, {Ade}, {Aghanim},
  {Arnaud}, {Ashdown}, {Aumont}, {Baccigalupi}, {Banday}, {Barreiro},
  {Bartlett} \& et al.}{{Planck Collaboration}
  et~al.}{2016}]{2016A&A...594A..13P}
{Planck Collaboration} {Ade} P.~A.~R.,  {Aghanim} N.,  {Arnaud} M.,  {Ashdown}
  M.,  {Aumont} J.,  {Baccigalupi} C.,  {Banday} A.~J.,  {Barreiro} R.~B.,
  {Bartlett} J.~G.,    et al. 2016, \aap, 594, A13

\bibitem[\protect\citeauthoryear{{Poulin}, {Lesgourgues} \& {Serpico}}{{Poulin}
  et~al.}{2017}]{2017JCAP...03..043P}
{Poulin} V.,  {Lesgourgues} J.,    {Serpico} P.~D.,  2017, \jcap, 3, 043

\bibitem[\protect\citeauthoryear{{Ruffini}, {Vereshchagin} \& {Xue}}{{Ruffini}
  et~al.}{2010}]{2010PhR...487....1R}
{Ruffini} R.,  {Vereshchagin} G.,    {Xue} S.-S.,  2010, \physrep, 487, 1

\bibitem[\protect\citeauthoryear{{Ruffini}, {Vereshchagin} \& {Xue}}{{Ruffini}
  et~al.}{2016}]{2016Ap&SS.361...82R}
{Ruffini} R.,  {Vereshchagin} G.~V.,    {Xue} S.-S.,  2016, \apss, 361, 82

\bibitem[\protect\citeauthoryear{{Stecker}, {de Jager} \& {Salamon}}{{Stecker}
  et~al.}{1992}]{1992ApJ...390L..49S}
{Stecker} F.~W.,  {de Jager} O.~C.,    {Salamon} M.~H.,  1992, \apjl, 390, L49

\bibitem[\protect\citeauthoryear{{Svensson} \& {Zdziarski}}{{Svensson} \&
  {Zdziarski}}{1990}]{1990ApJ...349..415S}
{Svensson} R.,  {Zdziarski} A.,  1990, \apj, 349, 415

\bibitem[\protect\citeauthoryear{{Zdziarski} \& {Svensson}}{{Zdziarski} \&
  {Svensson}}{1989}]{1989ApJ...344..551Z}
{Zdziarski} A.~A.,  {Svensson} R.,  1989, \apj, 344, 551

\end{thebibliography}

\end{document}